\documentclass[12pt]{article}
 \usepackage{epsfig}
 \def\be{\begin{equation}}
 \def\ee{\end{equation}}
 \def\bea{\begin{eqnarray}}
 \def\eea{\end{eqnarray}}
 \usepackage{graphicx}

 \catcode`\@=11
 \def\lsim{\mathrel{\mathpalette\@versim<}}
 \def\gsim{\mathrel{\mathpalette\@versim>}}
 \def\@versim#1#2{\vcenter{\offinterlineskip
 \ialign{$\m@th#1\hfil##\hfil$\crcr#2\crcr\sim\crcr } }}
 \catcode`\@=12

 \catcode`@=12
\begin{document}
 \thispagestyle{empty}
 \begin{flushright}
 UCRHEP-T592\\
 June 2018\
 \end{flushright}
 \vspace{0.6in}
 \begin{center}
 {\LARGE \bf Flavor Changing Neutral Currents in the\\ 
Asymmetric Left-Right Gauge Model\\}
 \vspace{1.0in}
 {\bf Chia-Feng Chang and Ernest Ma\\}
 \vspace{0.2in}
 {\sl Department of Physics and Astronomy,\\ 
 University of California, Riverside, California 92521, USA\\}
 \end{center}
 \vspace{1.0in}

\begin{abstract}\
In the $SU(3)_C \times SU(2)_L \times SU(2)_R \times U(1)_{(B-L)/2}$ extension 
of the standard model, a minimal (but asymmetric) scalar sector consists of 
one $SU(2)_R \times U(1)_{(B-L)/2}$ doublet and one $SU(2)_L \times SU(2)_R$ 
bidoublet.  Previous and recent studies have shown that this choice is 
useful for understanding neutrino mass as well as dark matter.  The 
constraints from flavor changing neutral currents mediated by the scalar 
sector are discussed in the context of the latest experimental data.
\end{abstract}

\newpage
 \baselineskip 24pt
\section{Introduction}
In the conventional left-right extension of the standard model (SM) of 
quarks and leptons, the gauge symmetry is 
$SU(3)_C \times SU(2)_L \times SU(2)_R \times U(1)_{(B-L)/2}$.  The scalar 
sector must be chosen to break $SU(2)_R \times U(1)_{(B-L)/2}$ to $U(1)_Y$ 
at a scale much higher than that of electroweak symmetry breaking, i.e. 
$SU(2)_L \times U(1)_Y$ to $U(1)_Q$.  This minimum requirement does not 
uniquely define the scalar particle content, i.e. doublets $\Phi_{L,R}$, 
triplets $\xi_{L,R}$, and bidoublets $\eta$.  There are basically 5 possible 
choices~\cite{m04} and they have implications on the nature of neutrino 
mass, as well as the $SU(2)_R$ breaking scale.  The simplest and often 
neglected choice is to have one $SU(2)_R \times U(1)_{(B-L)/2}$ doublet 
$\Phi_R$ and one $SU(2)_L \times SU(2)_R$ bidoublet $\eta$.  This implies 
by itself Dirac neutrino masses, but an inverse seesaw mechanism is easily 
implemented~\cite{admnw09} so that the observed neutrinos are Majorana 
fermions and the $SU(2)_R$ breaking scale is a few TeV.  Whereas 
flavor-changing neutral-current (FCNC) processes are unavoidable, they are 
manageable, as shown in Ref.~\cite{admnw09}.

Recently, it has been shown~\cite{m18} that such a model has another virtue, 
i.e. the appearance of predestined dark matter.  Because of the absence 
of an $SU(2)_L$ scalar doublet, the insertion of an $SU(2)_L$ fermion triplet 
$(\Sigma^+,\Sigma^0,\Sigma^-)$ or scalar triplet $(\chi^+,\chi^0,\chi^-)$ 
automatically guarantees either $\Sigma^0$ or $\chi^0$ to be stable, 
so that it is a good candidate for dark matter~\cite{ms09}.  Note that 
$\Sigma^0(\chi^0)$ is naturally lighter than $\Sigma^\pm(\chi^\pm)$ from 
radiative mass splitting~\cite{s95}.  A recently proposed model of 
$[SU(2)]^3$ dark matter~\cite{m18-1} also has this chosen scalar sector.

Since the writing of Ref.~\cite{admnw09}, there are new experimental 
results on FCNC, mostly in $B$ physics, and new theoretical calculations 
of their SM contributions.  In this paper, we update the resulting 
phenomenological contraints on this simple scalar sector consisting of 
only $\Phi_R$ and $\eta$.  In Sec.~2 the scalar sector is studied as 
well as the resulting massive gauge sector.  In Sec.~3 the Yukawa 
sector is studied and the structure of FCNC couplings to the physical 
neutral scalars is derived.  It is shown that under a simple assumption, 
all such effects depend only on two scalar masses which are almost degenerate 
in addition to an unknown unitary $3 \times 3$ matrix $V_R$ which is the 
right-handed analog of the well-known CKM matrix $V_{CKM}$ for left-handed 
quarks.  In Sec.~4 the experimental data on the $K-\bar{K}$, $B_d-\bar{B}_d$, 
and $B_s-\bar{B}_s$ mass differences, as well as the recent data on 
$B_s \to \mu^+ \mu^-$, are compared against their SM predictions to 
constrain the two scalar masses assuming that (A) $V_R = V_{CKM}$ and 
(B) $V_R = 1$.  In Sec.~5 there are some concluding remarks.

\section{Scalar and Gauge Sectors}
Under $SU(3)_C \times SU(2)_L \times SU(2)_R \times U(1)_{B-L}$, we assume 
one scalar doublet
\begin{equation}
\Phi_R = \pmatrix{\phi_R^+ \cr \phi_R^0} \sim (1,1,2,1/2)
\end{equation}
and one bidoublet
\begin{equation}
\eta = \pmatrix{\eta_1^0 & \eta_2^+ \cr \eta_1^- & \eta_2^0} \sim 
(1,2,,2,0).
\end{equation}
The dual of $\eta$, i.e. 
\begin{equation}
\tilde{\eta} = \sigma_2 \eta^* \sigma_2 = \pmatrix{\bar{\eta}_2^0 & 
-\eta_1^+ \cr -\eta_2^- & \bar{\eta}_2^0} \sim (1,2,2,0)
\end{equation}
is automatically generated and transforms exactly like $\eta$.

The most general Higgs potential consisting of $\Phi_R$, $\eta$, and 
$\tilde{\eta}$ is given by~\cite{admnw09}
\begin{eqnarray}
V &=& m_R^2 \Phi_R^\dagger \Phi_R + m^2 \hbox{Tr}(\eta^\dagger \eta) + {1 \over 2} 
\mu^2 \hbox{Tr}(\eta^\dagger \tilde{\eta} + \tilde{\eta}^\dagger \eta)  
+ {1 \over 2} \lambda_R (\Phi_R^\dagger \Phi_R)^2 + {1 \over 2} \lambda_1 
[\hbox{Tr}(\eta^\dagger \eta)]^2 \nonumber \\ &+& {1 \over 2} \lambda_2 
\hbox{Tr}(\eta^\dagger \eta 
\eta^\dagger \eta) + {1 \over 2} \lambda_3 \{ [\hbox{Tr}(\eta^\dagger 
\tilde{\eta})]^2 + [\hbox{Tr}(\tilde{\eta}^\dagger \eta)]^2 \} 
+ {1 \over 2} \lambda_4 \hbox{Tr}(\eta^\dagger \eta)[\hbox{Tr}(\eta^\dagger \tilde{\eta} + 
\tilde{\eta}^\dagger \eta)] \nonumber \\ &+& f_1 \Phi_R^\dagger 
(\tilde{\eta}^\dagger \tilde{\eta}) \Phi_R + f_2 \Phi_R^\dagger 
(\eta^\dagger \eta) \Phi_R + f_3 \Phi_R^\dagger (\eta^\dagger \tilde{\eta} 
+ \tilde{\eta}^\dagger \eta) \Phi_R,
\end{eqnarray}
where all parameters have been chosen real for simplicity.  Let 
$\langle \phi_R^0 \rangle = v_R$ and $\langle \eta^0_{1,2} \rangle = v_{1,2}$, 
then the minimum of $V$ has a solution where $v_2 \ll v_1$, i.e.
\begin{equation}
v_2 \simeq {-(\mu^2 + f_3 v_R^2 + \lambda_4 v_1^2)v_1 \over m^2 + f_2 v_R^2 + 
(\lambda_1 + \lambda_3)v_1^2},
\end{equation}
with
\begin{equation}
v_1^2 = {m_R^2 f_1 - m^2\lambda_R \over \lambda_R(\lambda_1+\lambda_2) - f_1^2}, 
~~~ v_R^2 = {-m_R^2 - f_1 v_1^2 \over \lambda_R}.
\end{equation}
In the limit $v_2=0$, the physical Higgs bosons are $\phi_2^\pm$ and 
$h_I = \sqrt{2}Im(\phi_2^0)$ with masses squared
\begin{equation}
m^2_\pm = (f_2-f_1)v_R^2, ~~~ m^2_I = (f_2-f_1)v_R^2 - (\lambda_2 + \lambda_3)v_1^2,
\end{equation}
and three linear combinations of $h_1 = \sqrt{2} Re(\phi_1^0)$, 
$h_2 = \sqrt{2} Re(\phi_2^0)$, and $h_R = \sqrt{2} Re(\phi_R^0)$, with 
the $3 \times 3$ mass-squared matrix
\begin{equation}
{\cal M}^2_h = \pmatrix{2(\lambda_1 + \lambda_2) v_1^2 & 2\lambda_4 v_1^2 & 
2f_1 v_1 v_R \cr 2\lambda_4 v_1^2 & (f_2-f_1)v_R^2 - (\lambda_2-\lambda_3)v_1^2 & 
2f_3 v_1 v_R \cr 2f_1 v_1 v_R & 2f_3 v_1v_R & 2\lambda_R v_R^2}.
\end{equation}
Since $v_1/v_R$ is known to be small, $h_{1,2,R}$ are approximately mass 
eignestates, with $h_1$ almost equal to the observed 125 GeV scalar boson 
at the Large Hadron Collider (LHC).  Note also that $h_2$ is almost 
degenerate with $h_I$ in mass.  We can make this even more precise 
by having small $\lambda_4$ and $f_{1,3}$.

There are two charged gauge bosons $W_L^\pm$ and $W_R^\pm$ in the 
$2 \times 2$ mass-squared matrix given by
\begin{equation}
{\cal M}^2_{W} = {1 \over 2}\pmatrix{g_L^2(v_1^2+v_2^2) & -2g_L g_Rv_1v_2 \cr 
-2g_L g_R v_1v_2 & g_R^2(v_R^2+v_1^2+v_2^2)}.
\end{equation}
With our assumption that $v_2 \ll v_1$, $W_L-W_R$ mixing is negligible. 
The present LHC bound on the $W_R$ mass is 3.7 TeV~\cite{WR18}.

There are three neutral gauge bosons, i.e. $W_{3L}$ from $SU(2)_L$, 
$W_{3R}$ from $SU(2)_R$, and $B$ from $U(1)_{(B-L)/2}$, with couplings 
$g_L$, $g_R$, and $g_B$ respectively.  Let them be rotated to the 
following three orthonormal states:
\begin{eqnarray}
A &=& {e \over g_L} W_{3L} + {e \over g_R} W_{3R}+ {e \over g_B} B, \\ 
Z &=& {e \over g_Y} W_{3L} - {e \over g_L} \left( {g_Y \over g_R} W_{3R} 
+ {g_Y \over g_B} B \right), \\ 
Z' &=& {g_Y \over g_B} W_{3R} - {g_Y \over g_R} B,
\end{eqnarray}
where
\begin{equation}
{1 \over e^2} = {1 \over g_L^2} + {1 \over g_Y^2}, ~~~ 
{1 \over g_Y^2} = {1 \over g_R^2} + {1 \over g_B^2}.
\end{equation}
The photon $A$ is massless and decouples from $Z$ and $Z'$, the latter 
two forming a mass-squared matrix given by
\begin{equation}
{\cal M}^2_Z = {1 \over 2}\pmatrix{(g_L^2+g_Y^2)(v_1^2+v_2^2) & 
-(g_L g_Y^2 g_R/eg_B)(v_1^2+v_2^2) \cr -(g_L g_Y^2 g_R/eg_B)(v_1^2+v_2^2) 
& (g_R^2+g_B^2)v_R^2 + (g_R^2 g_Y^2/g_B^2)(v_1^2+v_2^2)}.
\end{equation}
The neutral-current gauge interactions are given by
\begin{equation}
e A j_{em} + g_Z Z (j_{3L} - \sin^2 \theta_W j_{em}) + \sqrt{g_R^2+g_B^2} Z' 
\left[ j_{3R} + {g_Y^2 \over g_R^2}(j_{3L} - j_{em})\right].
\end{equation}
The present LHC bound on the $Z'$ mass is 4.1 TeV~\cite{ZP17}.  The $Z-Z'$ 
mixing is given by $(eg_R/g_B g_L) (m_Z^2/m_{Z'}^2)$ which is then 
less than $3.6 \times 10^{-4}$ for $g_R = g_L$ and within precision 
measurement bounds.

\section{Yukawa Sector and the FCNC Structure}
The fermion content is well-known, i.e.
\begin{eqnarray}
&& \psi_L = \pmatrix{\nu_e \cr e}_L \sim (1,2,1,-1/2), ~~~
\psi_R = \pmatrix{\nu_e \cr e}_R \sim (1,1,2,-1/2), \\ 
&&  q_L = \pmatrix{u \cr d}_L \sim (3,2,1,1/6), ~~~~~~~\;
q_R = \pmatrix{u \cr d}_R \sim (3,1,2,1/6),
\end{eqnarray}
with the electric charge given by $Q = I_{3L} + I_{3R} + (B-L)/2$.
Now the Yukawa couplings between the quarks and the neutral members of the 
scalar bidoublets are
\begin{equation}
(f^u_{ij} \eta_1^0 + f^d_{ij} \bar{\eta}_2^0) \bar{u}_{iL} u_{jR} + 
(f^u_{ij} \eta_2^0 + f^d_{ij} \bar{\eta}_1^0) \bar{d}_{iL} d_{jR}. 
\end{equation} 
In the limit $v_2=0$, both $up$ and $down$ quark masses come from only $v_1$. 
Hence
\begin{equation}
f^u_{ij}v_1 = U_L \pmatrix{m_u & 0 & 0 \cr 0 & m_c & 0 \cr 0 & 0 & m_t} 
U_R^\dagger, ~~~  
f^d_{ij}v_1 = D_L \pmatrix{m_d & 0 & 0 \cr 0 & m_s & 0 \cr 0 & 0 & m_b} 
D_R^\dagger,
\end{equation}
where $U_{L,R}$ and $D_{L,R}$ are unitary matrices, with
\begin{equation}
U_L^\dagger D_L = V_{CKM}, ~~~ U_R^\dagger D_R = V_R, 
\end{equation}
being the known quark mixing matrix for left-handed charged currents and 
the corresponding unknown one for their right-handed counterpart.

Whereas $Z$ and $Z'$ couple diagonally to all quarks, nondiagonal terms 
appear in the scalar Yukawa couplings.  Using Eqs.~(18), (19) and (20), 
the FCNC structure is then completely determined, i.e.
\begin{equation}
{h_1 \over \sqrt{2}v_1} \pmatrix{m_u & 0 & 0 \cr 0 & m_c & 0 \cr 0 & 0 & m_t} 
+ {(h_2 - i h_I) \over \sqrt{2}v_1} V_{CKM} \pmatrix{m_d & 0 & 0 \cr 0 & m_s 
& 0 \cr 0 & 0 & m_b} V_R^\dagger
\end{equation}
for the $up$ quarks, and
\begin{equation}
{h_1 \over \sqrt{2}v_1} \pmatrix{m_d & 0 & 0 \cr 0 & m_s & 0 \cr 
0 & 0 & m_b}  + {(h_2 + i h_I) \over \sqrt{2}v_1} V_{CKM}^\dagger 
\pmatrix{m_u & 0 & 0 \cr 0 & m_c & 0 \cr 0 & 0 & m_t} V_R
\end{equation}
for the $down$ quarks.  Hence $h_1$ behaves as the SM Higgs boson, and 
at tree-level, all FCNC effects come from $h_2$ and $h_I$.  We may thus 
use present data to constrain these two masses.  Note that all FCNC 
effects are suppressed by quark masses, so we have an understanding 
of why they are particularly small in light meson systems.

The analog of Eq.~(18) for leptons is
\begin{equation}
(f^\nu_{ij} \eta_1^0 + f^e_{ij} \bar{\eta}_2^0) \bar{\nu}_{iL} \nu_{jR} + 
(f^\nu_{ij} \eta_2^0 + f^e_{ij} \bar{\eta}_1^0) \bar{e}_{iL} e_{jR}. 
\end{equation} 
Hence
\begin{equation}
({\cal M}_\nu)_{ij} = f^\nu_{ij} v_1 + f^e_{ij} v_2, ~~~ 
({\cal M}_e)_{ij} = f^e_{ij} v_1 + f^\nu_{ij} v_2. 
\end{equation}
If neutrinos are Dirac fermions, then ${\cal M}_\nu \simeq 0$ compared to 
${\cal M}_e$, hence $f^\nu_{ij} = -(v_2/v_1) f^e_{ij}$ is a good 
approximation.  The analog of Eq.~(22) for charged leptons is then
\begin{equation}
\left[ {h_1 \over \sqrt{2}v_1} - {(h_2+ih_I) v_2 \over \sqrt{2} v_1^2} \right] 
\pmatrix{m_e & 0 & 0 \cr 0 & m_\mu & 0 \cr 0 & 0 & m_\tau}.
\end{equation}

\section{Phenomenological Constraints}
In the following we consider the contributions of Eqs.~(21), (22), and (25) to 
a number of processes sensitive to them in two scenarios: (A) $V_R = V_{CKM}$ 
and (B) $V_R = 1$.  We compare the most recent experimental data with 
theoretical SM calculations to obtain constraints coming from the mass 
differences $\Delta M_K$, $\Delta M_{B_d}$, $\Delta M_{B_s}$ of the neutral 
meson systems of $K-\bar{K}$, $B_d-\bar{B}_d$, $B_s-\bar{B}_s$ respectively, 
as well the recent measurement of~\cite{lhcb17} $B_s \to \mu^+ \mu^-$, i.e.
\begin{equation}
\bar{\mathcal{B}}\left( B_s \to \mu^+ \mu^- \right)_{\hbox{\scriptsize{LHCb}}} = 
\left( 3.0 \pm 0.6^{+0.3}_{-0.2}\right) \times 10^{-9},
\end{equation}
with an upper limit $\bar{\mathcal{B}} \left( B_d \to \mu^+ \mu^- 
\right)_{\hbox{\scriptsize{LHCb}}} < 3.4 \times 10^{-10}$ 
at $95\%$ confidence-level.  These values are in agreement 
with the next-to-leading-order (NLO) electroweak (EW) as well as 
NNLO QCD predictions \cite{bghmss14,usqcd16}:
\begin{equation}
\bar{\mathcal{B}}\left( B_s \to \mu^+ \mu^- \right)_{\hbox{\scriptsize{SM}}} = 
\left( 3.44 \pm 0.19\right) \times 10^{-9}, \;\;\;\;\;\;\;
\bar{\mathcal{B}}\left( B_d \to \mu^+ \mu^- \right)_{\hbox{\scriptsize{SM}}} = 
\left( 1.04 \pm 0.09\right) \times 10^{-10}.
\end{equation}
Nevertheless, new physics (NP) contributions are possible within the error 
bars. In addition, the $K$-$\bar{K}$ and $B_q$-$\bar{B}_q$ mixings, which 
interfere to obtain time-averaged decay widths 
\cite{dfn01,bfkkmt12,bfkkmpt12}, may also provide possible signals of NP. 

The most recently updated SM $\Delta M$ predictions 
\cite{usqcd16,flag17,swme16,clsz17,lkl18}, and the experimental 
measurements \cite{pdg16,hflav17} are
\begin{eqnarray}
\Delta M_K^{\hbox{\scriptsize{exp}}} &= \left( 5.296 \pm 0.009 \right)  
\hbox{fs}^{-1}, \;\;\;\;\;\;\;\;\;\;\;\;\;\;\;\;\;\;\;\, 
\Delta M_K^{\hbox{\scriptsize{SM}}} = \left( 4.73 \pm 1.91 \right)  
\hbox{fs}^{-1},\\
\Delta M_{B_d}^{\hbox{\scriptsize{exp}}} &= \left( 0.5055 \pm 0.0020 \right) 
\hbox{ps}^{-1}, \;\;\;\;\;\;\;\;\;\;\;\;\;\;\; 
\Delta M_{B_d}^{\hbox{\scriptsize{SM}}} 
= \left( 0.642 \pm 0.069 \right) \hbox{ps}^{-1},\\
\Delta M_{B_s}^{\hbox{\scriptsize{exp}}} &= \left( 17.757 \pm 0.021 \right) 
\hbox{ps}^{-1}, \;\;\;\;\;\;\;\;\;\;\;\;\;\;\;\;\, 
\Delta M_{B_s}^{\hbox{\scriptsize{SM}}} = \left( 20.01 \pm 1.25 \right) 
\hbox{ps}^{-1}.
\end{eqnarray}
Note that $\Delta M_{B_d}^{\hbox{\scriptsize{SM}}}$ is estimated by the 
$SU(3)$-breaking ratio $\xi = 1.206(18)(6)$ \cite{usqcd16}, and the 
NLO EW, NNLO QCD corrections have been incorporated as well.

\subsection{$\Delta M_{B_q}$ and $\Delta M_{K}$}
In the SM, other than long-distance contributions~\cite{clsz17}, 
$B_q-\bar{B}_q$ and $K-\bar{K}$ mixings occur mainly via the well-known box 
diagrams with the exchange of $W^{\pm}$ bosons and the $(u,c,t)$ quarks.  In the 
asymmetric left-right model, the new scalars $h_2$ and $h_I$ have additional 
tree-level contributions.  We consider the usual operator analysis with Wilson 
coefficients obtained from the renormalization group (RG).  The mass 
difference between the two mass eigenstates of a neutral meson system 
(see \cite{pdg16,abl16} for details) may be obtained from the $\Delta F = 2$ 
effective Hamiltonian \cite{b98,bju01,chyy17}
\begin{equation}
\mathcal{H}_{\hbox{\small{eff}}}^{\Delta F = 2} = \frac{G_F^2}{16\pi^2} m_W^2 
\left( V_{tb} V_{tq}^*\right)^2 \sum_i C_i \mathcal{O}_i + \hbox{H.c.},
\end{equation}
where the operators relevant to the SM and the new scalar contributions 
are \cite{usqcd16} 
\begin{eqnarray}
\mathcal{O}_{SM} = & \left(\bar{b}^\alpha \gamma_\mu P_L q^\alpha \right)
\left(\bar{b}^\beta \gamma_\mu P_L q^\alpha \right), \;\;\;\;\; \mathcal{O}_4 
= \left( \bar{b}^\alpha P_L q^\alpha \right) \left( \bar{b}^\beta P_R q^\beta 
\right),\\
\mathcal{O}_2 = & \left( \bar{b}^\alpha P_L q^\alpha \right) \left( 
\bar{b}^\beta P_L q^\beta \right), \;\;\;\;\;\;\;\;\;\;\;\; 
\tilde{\mathcal{O}}_2 = \left( \bar{b}^\alpha P_R q^\alpha \right) \left( 
\bar{b}^\beta P_R q^\beta \right),\\
\mathcal{O}_3 = & \left( \bar{b}^\alpha P_L q^\beta \right) \left( 
\bar{b}^\beta P_L q^\alpha \right), \;\;\;\;\;\;\;\;\;\;\;\; 
\tilde{\mathcal{O}}_3 = \left( \bar{b}^\alpha P_R q^\beta \right) \left( 
\bar{b}^\beta P_R q^\alpha \right),
\end{eqnarray}
for the $B_d - \bar{B}_d$ and $B_s - \bar{B}_s$ systems.  In the case of 
$K - \bar{K}$, we just change $b$ to $s$ and $q$ to $d$ in the above. 
$P_R$ and $P_L$ are right- and left-handed projection operators 
$(1\pm \gamma_5)/2$, respectively.  $\alpha$ and $\beta$ are color indices. 
We follow the details in \cite{b98} with recent updates \cite{usqcd16,llp14} 
for $B_q$ as well as \cite{clsz17} for $K$.  After ignoring terms that are 
suppressed by light quark masses, we obtain
\begin{equation}
C_{SM}^q = 4 S_0 (x_t) \eta_{2B}(\mu), \;\;\;\;\; C_{SM}^K = 4 \lambda_c^2 
\eta_{cc} S_0 (x_c)/\lambda_t^2 + 4 \eta_{tt} S_0 (x_t) + 8 \lambda_c \eta_{ct} 
S_0 (x_c,x_t)/\lambda_t,
\end{equation}
with $\lambda_x \equiv V_{xs} V^*_{xd}$.  The Inami-Lim function $S_0(x_i,x_j)$ 
with $x_q \equiv \left( m_q(m_q)/m_W\right)^2$ describes the electroweak 
corrections in one loop \cite{il81}.  The factors $\eta_{i}$ are perturbative 
QCD corrections at NLO \cite{b98}, as well as \cite{bmz97}(\cite{bju01}) 
for the new $B_q(K)$ terms.  Since the QCD corrections generate nondiagonal 
entries, the color mixed operators should be considered as well at low scale 
\cite{bbgh12} (see also \cite{flag17,bmu00,gjnt11}).

Noting that 
$\langle \mathcal{O}_{2,3} \rangle =\langle \tilde{\mathcal{O}}_{2,3} \rangle$ 
in QCD, we consider the relevant operators for $B_q - \bar{B}_q$ mixing 
in terms of their bag parameters \cite{usqcd16,ggms96},
\begin{equation}
\label{OM1}
\langle \mathcal{O}_1^q \rangle (\mu) = c_1 f_{B_q}^2 M_{B_q}^2 B^{(1)}_{B_{q}} (\mu)
\end{equation}
and 
\begin{equation}
\label{OM2}
\langle \mathcal{O}_i^q \rangle (\mu) = c_i \left( \frac{M_{B_q}}{m_b(\mu) 
+ m_q} \right)^2 f^2_{B_q} M^2_{B_q} B^{(i)}_{B_q} (\mu), \;\;\;\;\;\;\;\;\;\;\;\; 
i = 2,3,
\end{equation}
and
\begin{equation}
\label{OM3}
\langle \mathcal{O}_i^q \rangle (\mu) = c_i \left[ \left( 
\frac{M_{B_q}}{m_b(\mu) + m_q} \right)^2 + d_i \right] f^2_{B_q} M^2_{B_q} 
B^{(i)}_{B_q} (\mu), \;\;\;\;\;\;\;\;\;\;\;\; i = 4,5,
\end{equation}
with
$c_i = \{2/3,-5/12,1/12,1/2,1/6 \}$, $d_4 = 1/6$, and $d_5 = 3/2$. The decay 
constants and bag parameters $B^{(i)}_{B_q}$ include all nonperturbative 
effects.  The lattice calculation has been done in \cite{usqcd16} for $B_q$ 
with in the scheme of \cite{bmu00}, as well as \cite{swme16} for $K$. 
The renormalization group evolution effects are considered in 
\cite{bju01,bmz97}.

In the asymmetric left-right model, the tree-level $h_2$ and $h_I$ 
contributions to the Wilson coefficients at the new physics scale 
$\mu_{\hbox{\scriptsize{NP}}}$ are
\begin{equation}
\label{Wilson1}
C_2 = - \frac{1}{2} \kappa \left[ \left( V_d^\dagger \right)_{b,q} \right]^2  
\left( \frac{1}{m_2^2} - \frac{1}{m_I^2} \right), \;\;\;
\tilde{C}_2 = - \frac{1}{2} \kappa \left[ \left( V_d \right)_{b,q} \right]^2 
\left( \frac{1}{m_2^2} - \frac{1}{m_I^2} \right),
\end{equation}
\begin{equation}
\label{Wilson2}
C_4 = - \kappa \left( V_d \right)_{b,q} \left( V_d^\dagger \right)_{b,q} 
\left( \frac{1}{m_2^2} + \frac{1}{m_I^2} \right),
\end{equation}
where $\kappa = 16 \pi^2/G_F^2 m_W^2 \left( V_{tb} V_{tq}^* \right)^2$, 
and the matrix $V_d$ comes from the second term of Eq.(22).  The $B_q$ mass 
difference is thus given by
\begin{eqnarray}
&& 2 M_{12}^q = {\langle \bar{B}_q | \mathcal{H}_{\hbox{\small{eff}}}^{\Delta F = 2} 
| B_q \rangle \over M_{B_q}} = \frac{G_F^2}{16\pi^2} \frac{m_W^2}{M_{B_q}} 
\left( V_{tb} V_{tq}^* \right)^2 \times \nonumber \\ 
&& \left[  C_{SM}^q c_1 f_{B_q}^2 M_{B_q}^2 \hat{B}^{(1)}_{B_q}+
\left( C_2 + \tilde{C}_2\right) \left( \eta_{22} \langle \mathcal{O}_2^q 
\rangle + \eta_{32} \langle \mathcal{O}_3^q \rangle \right) + C_4 \eta_4 
\langle \mathcal{O}_4^q \rangle \right],
\end{eqnarray}
where $\eta_4 \simeq 3.90$, $\eta_{22} \simeq 2.25$ and 
$\eta_{32} \simeq -0.12$, \cite{bmz97,ghpp07}. Similarly, the 
$K^0$ mass difference is 
\begin{eqnarray}
&& 2 M_{12}^K = \frac{G_F^2 m_W^2}{16\pi^2} f_K^2 M_K
\left( V_{ts} V_{td}^* \right)^2 \left[  C_{SM}^K P_1^{VLL}  +
 C_2 P_1^{SLL} + C_4 P_2^{LR}  \right],
\end{eqnarray}
where $P_{1,2}$ are given in \cite{clsz17,bju01} and a recently updated 
lattice simulation \cite{swme16}. Hence 
\begin{equation}
\Delta M_{K} = 2 \hbox{Re}\left[ M_{12}^K \right], \;\;\;\;\;  
\Delta M_{B_q} = 2 \left| M_{12}^q \right|, \;\;\;\;\; \hbox{and} \;\;\;\;\; 
\phi_q^M = \hbox{arg} M_{12}^q.
\end{equation}
Note that $\phi^M_s$ may deviate \cite{bfkkmpt12} from the SM value, i.e. 
$\phi_s^M = \phi_s^{SM} + \phi^{NP}_s$.  A nonzero $\phi^{NP}_s$ would 
contribute to the $CP$ violation effect in the $B_{s} \to (J/\psi) \phi$ 
decay (see \cite{bfgk13} and the recent review \cite{abl16}).  Present data 
imply the constraint $\phi^{NP}_s = 0.4^\circ \pm 1.9^\circ$ \cite{fgjt18}.  
For $B_d$, the phase constraint is 
$\phi_d^{NP} = -3.8^\circ \pm 4.4^\circ$ \cite{ckm15,agrt18}.

\subsection{$B_s \to \mu^+ \mu^-$}
The scalars $h_2$ and $h_I$ contribute not only to the mass difference of 
$B_{s}$, but also to the decay of $B_s \to \mu^+ \mu^-$ at tree level. 
The SM contribution is dominated by the operator $\mathcal{O}_{10}^{SM}$, so 
we ignore other possible SM operators \cite{chyy17,llp14}.  The effective 
Hamiltonian is given by \cite{bghmss14,bbl96}
\begin{equation}
\mathcal{H}_{\hbox{\small{eff}}} = - \frac{G_F}{\sqrt{2}} \frac{\alpha_{em}}
{\pi s_W^2} V_{tb} V_{ts}^* \left( C_{10}^{SM}\mathcal{O}_{10}^{SM} + C_S 
\mathcal{O}_S + C_P \mathcal{O}_P + C_S^\prime \mathcal{O}_S^\prime + 
C_P^\prime \mathcal{O}_P^\prime \right) + \hbox{H.c.},
\end{equation}
where $\alpha_{em}$ is the fine structure constant, and 
$s_W^2 \equiv \sin^2\theta_W$ with $\theta_W$ the weak mixing angle. 
The operators are defined as
\begin{equation}
\mathcal{O}_{10}^{SM} =  \left( \bar{q} \gamma^\mu P_L b  \right)  
\left( \bar{\mu} \gamma^\mu \gamma_5 \mu \right), \;\;\;\;\; 
\mathcal{O}_P = m_b \left( \bar{q} P_R b \right) 
\left( \bar{\mu} \gamma_5 \mu\right), \;\;\;\;\; 
\mathcal{O}_P^\prime = m_b \left( \bar{q} P_L b\right) 
\left( \bar{\mu} \gamma_5 \mu\right),
\end{equation}
\begin{equation}
\mathcal{O}_S = m_b \left( \bar{q} P_R b \right) \left( \bar{\mu}\mu \right), 
\;\;\;\;\; \mathcal{O}_S^\prime = m_b \left( \bar{q} P_L b \right) 
\left( \bar{\mu}\mu \right).
\end{equation}
Including the $b$ quark mass $m_b$ makes those operators as well as their 
Wilson coefficients to be renormalization-group invariant \cite{llp14}. 
For the NLO SM contribution, we use a numerical value approximated 
by \cite{llp14}
\begin{equation}
C_{10}^{SM} = -0.9380 \left( \frac{m_t^p}{173.1\hbox{ GeV}} \right)^{1.53} 
\left( \frac{\alpha_s \left( m_Z \right)}{0.1184} \right)^{-0.09},
\end{equation}
where $m_t^p$ is the t quark pole mass. The contributions of NLO EW and 
NNLO QCD have been computed by \cite{bb93,mu99,bb99,bgs14,hms13}. 
The non-SM Wilson coefficients are given by tree-level $h_2$ or $h_I$ 
exchange, i.e.
\begin{equation}
\label{Wilson3}
C_P = \tilde{\kappa} \left( V_d \right)_{s,b} \left( i\frac{\hbox{Im} 
\left( V_p \right)_{\mu\mu}}{m_2^2} - \frac{\hbox{Re}\left( V_p 
\right)_{\mu\mu}}{m_I^2} \right), \;\;\;
C_P^\prime = \tilde{\kappa} \left( V_d^\dagger \right)_{s,b} \left( 
i\frac{\hbox{Im} \left( V_p \right)_{\mu\mu}}{m_2^2} + 
\frac{\hbox{Re}\left( V_p \right)_{\mu\mu}}{m_I^2} \right),
\end{equation}
\begin{equation}
C_S = \tilde{\kappa} \left( V_d \right)_{s,b} \left( \frac{\hbox{Re} 
\left( V_p \right)_{\mu\mu}}{m_2^2} - i\frac{\hbox{Im}\left( V_p 
\right)_{\mu\mu}}{m_I^2} \right), \;\;\;
C_S^\prime = \tilde{\kappa} \left( V_d^\dagger \right)_{s,b} 
\left( \frac{\hbox{Re} \left( V_p \right)_{\mu\mu}}{m_2^2} + 
i\frac{\hbox{Im}\left( V_p \right)_{\mu\mu}}{m_I^2} \right),
\end{equation}
where $\tilde{\kappa} = \pi^2/G_F^2 m_b m_W^2 V_{tb} V_{ts}^*$, and the matrix 
$V_p$ comes from the second term of Eq.~(25). The form factors are
\begin{equation}
\langle 0 | \bar{q} \gamma_\mu \gamma_5 b | \bar{B}_{q} (p) \rangle = 
i f_{B_q} p_\mu, \;\;\;\;\; (m_b) \langle 0 | \bar{q} \gamma_5 b | 
\bar{B}_q (p) \rangle = - i f_{B_q} \frac{M^2_{B_q}}{m_b + m_q} m_b.
\end{equation}
From the above, the branching fraction of $B_s \to \mu^+ \mu^-$ is 
then \cite{dfn01}
\begin{equation}
\mathcal{B} \left( B_q \to \mu^+ \mu^- \right) = \frac{ \tau_{B_q} G_F^4 m_W^4}
{ 8 \pi^5} \left| V_{tb} V_{ts}^* \right|^2 f_B^2 m_\mu^2 m_B 
\sqrt{1 - \frac{4 m_\mu^2}{m_B^2}} \left( |P|^2 + |S|^2 \right),
\end{equation} 
where $m_{B_s}$, $\tau_{B_s}$ and $f_{B_s}$ denote the mass, lifetime and decay 
constant of the $B_s$ meson, respectively. The amplitudes $P$ and $S$ are 
defined as \cite{bfkkmpt12}
\begin{equation}
P \equiv  C_{10}^{SM} + \frac{m_B^2 m_b}{2 m_\mu \left(m_b + m_q \right)} 
\left( C_P - C_P^\prime \right), \;\;\;\;\;\;\;\;
S \equiv  \sqrt{1-\frac{4 m_\mu^2}{m_B^2}} \frac{m_B^2 m_b}{2m_\mu 
\left(m_b + m_q \right)} \left( C_S - C_S^\prime \right).
\end{equation}
To compare against experimental data, the time-integrated branching fraction 
is discussed extensively in \cite{dfn01,bfkkmt12,bfkkmpt12,dr86}, i.e.
\begin{equation}
\bar{\mathcal{B}}\left( B_s \to \mu^+ \mu^- \right)_{\hbox{\small{exp}}} = 
\left( \frac{1 + \mathcal{A}_{\Delta \Gamma} y_s}{1-y_s^2} \right) 
\mathcal{B}\left( B_s \to \mu^+ \mu^- \right),
\end{equation}
where $y_s = \Delta \Gamma_s/2\Gamma_s$ ($\Gamma_s$ being the average 
$B_s$ decay width) and \cite{bfgk13}
\begin{equation}
\mathcal{A}_{\Delta \Gamma} = \frac{|P|^2 \cos \left( 2 \phi_P - \phi_s^{NP} 
\right) - |S|^2 \cos \left( 2 \phi_S - \phi_s^{NP} \right)}{|P|^2+|S|^2},
\end{equation}
with
\begin{equation}
S = |S| e^{i\phi_S}, \;\;\;\;\; P = |P| e^{i\phi_P}, \;\;\;\;\;2 \hbox{arg} 
\left( V_{ts} V^*_{tb} \right) \equiv \phi_s^{SM}.
\end{equation}

\subsection{Numerical Analysis}
We now discuss the experimental constraints on the two scalar 
masses $m_2$ and $m_I$.  We allow for the theoretical uncertainties in 
computing $\Delta M_{K}$, $\Delta M_{B_q}$ and $B_s \to \mu^+ \mu^-$ which 
arise mainly from the decay constant $f_{B_q}$ (and the bag parameters 
$\hat{B}_q^{(i)}$) and the combination of CKM matrix elements 
$|V_{ts}^* V_{tb}|$ 
(i.e. $|V_{cb}|$ as well as $|V_{ub}|$, from the unitarity of $V_{CKM}$) 
\cite{usqcd16}. We note that there is a long-standing discrepancy between 
the determinations of $V_{ub}$ from inclusive and exclusive $B$ decays.  
We adopt the recent averaged CKM matrix elements by the CKMfitter 
group \cite{ckm15}, and use running quark masses \cite{xzz12}. 
Our input parameters are given in Table \ref{table1}, and the scales used 
are $\{\mu_K,\mu_b,\mu_{\hbox{\scriptsize{NP}}}\} = \{2,3,1000\}$ GeV.
\begin{table}[b!]
\caption{List of input parameters (including Table XIII of \cite{usqcd16} 
in the scheme of \cite{bmu00}).} 
\centering 
\begin{tabular}{| c | c | c | c | c | c |} 
\hline 
Parameter & Value & Ref. & Parameter & Value & Ref. \\  
\hline\hline 
$m_W$ & $80.385(15)$ GeV & \cite{pdg16} & $m_t^p$ & $\;\;\; 173.21(87)$ GeV 
$\;\;\;$ & \cite{pdg16} \\ \hline 
$G_F$ & $\; 1.1663787(6)\times 10^{-5}$ GeV$^{-2} \;$ & \cite{pdg16} & 
$m_t(m_t)$ & $162.5(11)$ GeV & \cite{xzz12} \\ \hline
$\hbar$ & $\;\; 6.582119514(40) \times 10^{-25}$ GeV s $\;$ & \cite{pdg16} & 
$m_b(m_b)$ & $4.19(18)$ GeV & \cite{xzz12}  \\ \hline
$\tau_{B_s}$ & $1.510(5)$ ps & \cite{pdg16} & $m_c(m_b)$ & 
$0.934^{+0.058}_{-0.120}$ GeV & \cite{xzz12} \\ \hline
$\Delta \Gamma_{s}$ & $0.082(7)$ ps$^{-1}$ & \cite{pdg16} & 
$m_s(m_b)$ & $84^{+26}_{-17}$ MeV & \cite{xzz12} \\ \hline
$M_{B_s}$ & $5.36689(19)$ GeV & \cite{pdg16} & $m_u(m_b)$ & $2.02(60)$ MeV & 
\cite{xzz12} \\ \hline
$M_{B_d}$ & $5.27961(16)$ GeV & \cite{pdg16} & $m_d(m_b)$ & $4.12(69)$ MeV 
& \cite{xzz12} \\ \hline
 $M_K$ & $0.497611(13)$ GeV & \cite{pdg16} & $m_c (m_c)$ & 
$1.29^{+0.05}_{-0.11}$ GeV & \cite{xzz12} \\ \hline
$\alpha_{s}^{(5)}\left( m_Z \right)$ & $0.1181(11)$ & \cite{pdg16} & 
$V_{us}$ & $0.22508^{+0.00030}_{-0.00028}$  & \cite{ckm15} \\ \hline
$f_{B_s}$& $227.2(34)$ MeV  & \cite{usqcd16} &$V_{cb}$ & 
$0.04181^{+0.00028}_{-0.00060}$  & \cite{ckm15} \\ \hline
$\gamma_{\hbox{\tiny{CKM}}}$ & $1.141^{+0.017}_{-0.020}$  & \cite{ckm15} & 
$|V_{ub}/V_{cb}|$ & $0.0889(14)$ & \cite{ckm15} \\ \hline
$f_K$ & $0.1562(9)$ GeV  & \cite{flag17} & $\hat{B}_K$ & $0.7625(97)$ 
& \cite{flag17} \\ \hline
$B_{K}^{(2)}(2\;\hbox{GeV})$ & $0.568(26)$  & \cite{swme16} & 
$B_{K}^{(3)}(2\;\hbox{GeV})$ & $0.382(21)$ & \cite{swme16} \\ \hline
$B_{K}^{(4)}(2\;\hbox{GeV})$ & $0.984(67)$  & \cite{swme16} & 
$B_{K}^{(5)}(2\;\hbox{GeV})$ & $0.714(78)$ & \cite{swme16} \\ \hline
$f_{B_d}$ & $190.9(4.1)$ MeV & \cite{usqcd16} & $\eta_{cc}$ & $1.87(76)$ 
& \cite{bbcj17} \\ \hline
  $\Lambda^{(5)}_{\hbox{\scriptsize{QCD}}}$ & $0.226$ GeV & \cite{b98} & 
$\eta_{ct}$ & $0.496(47)$ & \cite{bbcj17} \\ \hline
$f_{B_s}\sqrt{\hat{B}^{(1)}_{B_s}}$ & $274.6 \pm 11.1 $ MeV & \cite{usqcd16} 
& $\eta_{tt}$ & $0.5765(65)$ & \cite{bbcj17} \\ \hline
    $f_{B_d}\sqrt{\hat{B}^{(1)}_{B_d}}$ & $227.7 \pm 11.8 $ MeV 
& \cite{usqcd16} & $\eta_{2B}$ & $0.55210(62)$ & \cite{bbl96}\\ \hline
\end{tabular}
\label{table1} 
\end{table}

\begin{figure}[t!]
\centering
\includegraphics[width=0.6\textwidth]{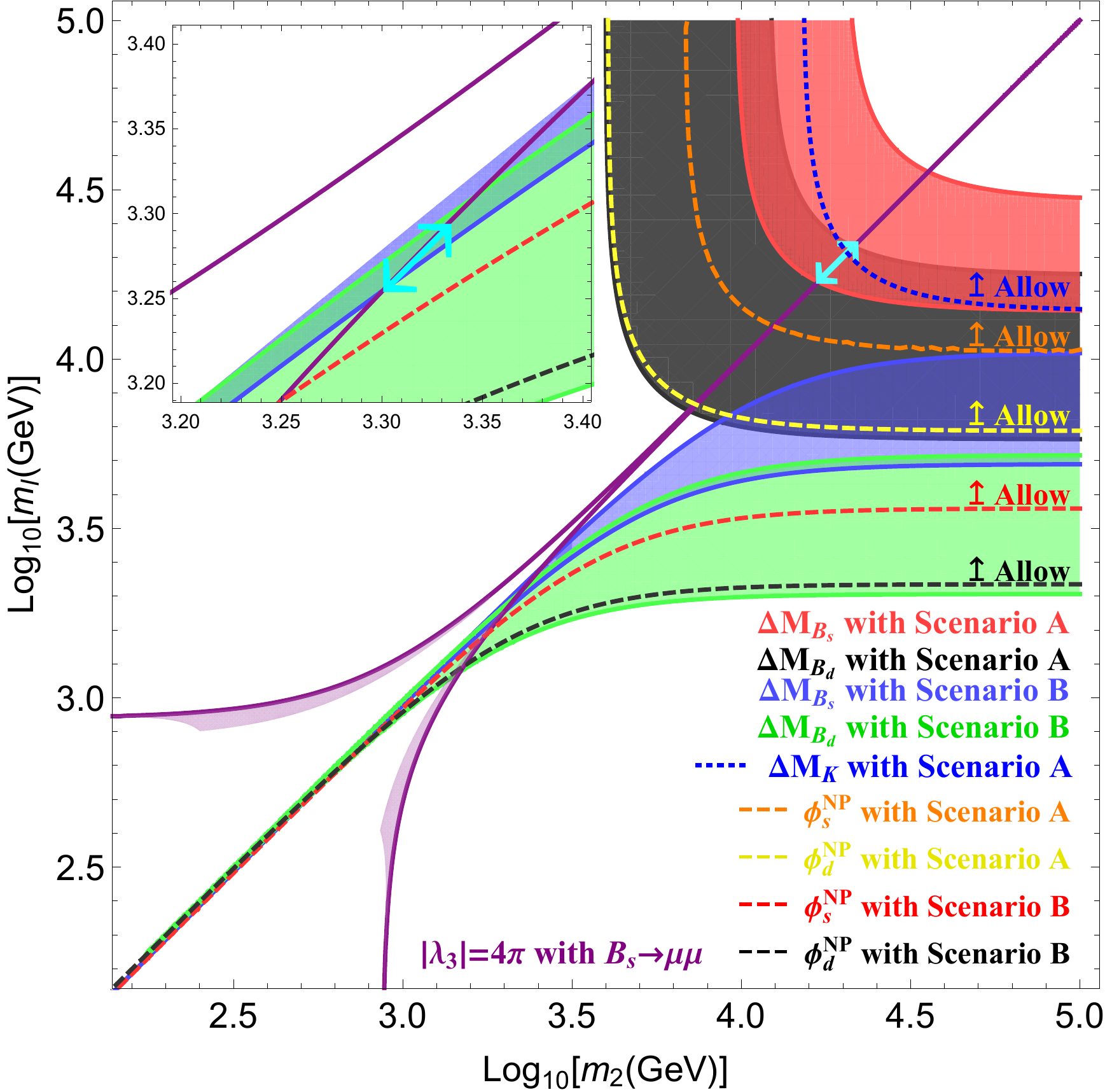} 
\caption{ \label{FFCNC} \baselineskip 16.5pt 
Allowed parameter space in the $\left(m_2, m_I\right)$ plane. The (red, black) 
and (blue, green) shaded regions correspond to Scenario A ($V_R = V_{CKM}$) 
and Scenario B ($V_R = 1$) within the $1\sigma$ region of $\Delta M_{B_q}$, 
respectively.  The purple shaded regions correspond to Scenario (A, B) 
with $v_2< \frac{1}{2} v_1$ from the constraint 
$m_2^2-m_I^2 = 2 \lambda_3 v_1^2$ with $|\lambda_3| = 4\pi$ and its 
overlap within the $1\sigma$ region of 
$\bar{\mathcal{B}}(B_s \to \mu^+ \mu^-)$. The dotted blue line corresponds 
to the $\Delta M_K$ constraint, including the LD effects. The 
light-orange(dashed yellow line) is shown at the $1\sigma$ experimental 
CP phase constraint of the 
$B_s(B_d)$ phase in Scenario A, and the dashed red(black) line is the 
$B_s(B_d)$ phase constraint in Scenario B, which excludes the lower-right 
region of this figure. The dark-purple lines show the $v_2 \to 0$ limit, 
i.e. a null contribution to $B_s \to \mu \mu$ from new physics.  The
survival parameter spaces under $\Delta M_{s,d}$ are marked by cyan 
$\updownarrow$.  The input parameters are from Table \ref{table1}.}
\end{figure}
Flavor-changing neutral scalar couplings to quarks are studied in two 
scenarios, where the $SU(2)_R$ charged-current mixing matrix 
$V_R$ in Eq.(20) is given either by the CKM matrix (Scenario A), 
i.e. $V_R \equiv V_{CKM}$, or just the identity matrix (Scenario B), 
i.e. $V_R \equiv 1$.  Tree-level contributions exist from the 
exchange of the new CP-even scalar $h_2$ or the CP-odd scalar $h_I$,  
as shown in Fig. \ref{FFCNC}.  The Wilson 
coefficients for $\Delta M_{B_q}$ and $\Delta M_K$ are given in 
Eqs.(39) and (40).  The $B_s \to \mu^+\mu^-$ 
contribution comes from Eqs.(47) and (48).

In Scenario A, since the mixing matrix $V_d$ is Hermitian 
[see Eq.(22)], fine-tuned cancellations between $C_2$, $\tilde{C}_2$ 
and $C_4$ appear only if a large ratio 
$\left( m_2^2 - m_I^2\right)/ \left( m_2^2 + m_I^2\right)$ appears, 
[see Eq.(41)], but this cannot happen within the given parameter space.
Therefore, the $\Delta M_{B_q}$ constraints only allow the (red, black) 
area without fine-tuning, i.e. $m_2$ and/or $m_I \geq 13.5 \;\hbox{TeV}$.  
On the other hand, the $h_2-h_I$ mass-squared difference 
$m_2^2 - m_I^2 = 2 \lambda_3 v_1^2$ restricts it to only a thin line 
in the region of heavier masses, i.e. $m_2 \simeq m_I$.  Their overlap 
shows a strong constraint indicated by an arrow (cyan) in Fig.~1.  
If the $\Delta M_K$ constraint is included, then this tiny allowed 
region is ruled out if only the short-distance (SD) contribution is 
considered.  Adding the long-distance (LD) contributions from $\pi$ and 
$\eta'$ exchange \cite{bgi10,gst05}
\begin{equation}
\Delta m_K = \Delta m_K^{SD} + \Delta m_K^{LD}|_{\pi\pi} + 
\Delta m_K^{LD}|_{\eta^\prime},
\end{equation}
with
\begin{equation}
\Delta m_K^{LD}|_{\pi\pi} = 0.4 \Delta m_{K}^{\hbox{\scriptsize{exp}}}, 
\;\;\;\;\;\;
\Delta m_K^{LD}|_{\eta^{\prime}} = -0.3 \Delta m_{K}^{\hbox{\scriptsize{exp}}},
\end{equation}
a consistent overlap with the data may be obtained. 
Although the LD contributions are still not well understood, 
with somewhat large uncertainties \cite{clsz17},  
these terms shift the SM contribution and allow Scenario A to survive. 
In summary, the above constraints with LD physics allow the masses 
to lie within the region 
$20.0 \; \hbox{TeV} \leq m_2 \simeq m_I \leq 22.8\; \hbox{TeV}$.

In Scenario B, the asymmetric mixing matrix elements e.g. 
$\left(V_d\right)_{b,s} \simeq - 0.01(V_d^\dagger)_{b,s}$ 
result in cancellations between Wilson coefficients $C_2$, $\tilde{C_2}$ 
and $C_4$ if 
$ \left( m_2^2 - m_I^2\right)/ \left( m_2^2 + m_I^2\right) \simeq 0.01$. 
Hence lighter $m_2$, $m_I$ masses from $\Delta M_{B_q}$ are not ruled out 
in the (blue, green) area of Fig.~\ref{FFCNC} where $|\lambda_3| = 4\pi$ 
has been used.  The two branches (purple) represent the model restrictions 
on $(m_2,m_I)$ depending on the sign of $\lambda_3$.  If a value of 
$|\lambda_3|$ less than $4\pi$ is used, then the region between these 
two branches will be filled in.  Since our model contribution to 
$B_s \to \mu^+\mu^-$ is proportional to $v_2$ which is always assumed 
to be small so far, there is no constraint from it unless $v_2$ is 
sizeable.  For $|\lambda_3|=4\pi$, if we also assume $v_2 < 0.5 v_1$, 
then within 1$\sigma$ of the $B_s \to \mu^+\mu^-$ experimental rate, 
the allowed region cuts off for small $(m_2,m_I)$, as shown (purple) 
in Fig.~1.  The allowed region with $\lambda_3=4\pi$ in Scenario B is 
indicated by an arrow (cyan) in the subgraph, i.e. 
$1.80\leq m_I \leq2.45$ TeV.  For $\lambda_3 < 4\pi$, a thin region opens 
up above the purple line.   As for $\Delta M_{K}$ in Scenario B, 
this result is not affected whether LD contributions are included or not. 

From Eq.(21), we see that $D^0-\bar{D}^0$ mixing is suppressed by 
down-quark masses in the asymmetric left-right model.  It does not 
provide a tighter constraint \cite{ghpp07,usqcd18,bc18}.

\section{Concluding Remarks}
We have studied the possible contributions of the heavy scalars $h_2$ and 
$h_I$ in the asymmetric left-right model to $B_q - \bar{B}_q$ mixings 
as well as $B_s \to \mu^+ \mu^-$.  We find that improvements of the fit 
to experimental data within $1\sigma$ are possible, as shown in Fig.~1. 
In the scenario with the right-handed charged-current mixing matrix $V_R$ 
equal to $V_{CKM}$, we predict $m_2 \simeq m_I$ to be between 20.0 and 22.8 
TeV. If $V_R = 1$, then $m_I \simeq 1.80$ to 2.45 TeV, and $m_2 \simeq 2.00$ 
to 2.60 TeV for $\lambda_3 = 4\pi$ and small $v_2$.

If the doublet $\Phi_R$ is replaced with the triplet 
$(\xi_R^{++},\xi_R^+,\xi_R^0)$, the FCNC analysis remains the same. 
What will change is that $\nu_R$ will acquire a large Majorana mass and 
the usual neutrinos will get seesaw Majorana masses.  A doubly-charged 
physical scalar $\xi_R^{\pm \pm}$ will also appear and decays to 
$e^\pm e^\pm$. 
In addition, there are more candidates for predestined dark matter~\cite{m18}, 
i.e. scalar $SU(2)_L$ triplet, fermion singlet, fermion bidoublet, fermion 
$SU(2)_L$ triplet, and fermion $SU(2)_R$ triplet.

\section*{Acknowledgement}
This work was supported in part by the U.~S.~Department of Energy Grant 
No. DE-SC0008541.

\baselineskip 18pt
\bibliographystyle{unsrt}

\end{document}